\documentclass[aps,twocolumn,groupedaddress]{revtex4-1}

\usepackage{amsmath}
\usepackage{mathtools}
\usepackage{array}
\usepackage{gensymb}
\usepackage{multirow}
\usepackage[utf8]{inputenc}
\usepackage{booktabs}
\usepackage{longtable}
\usepackage{tabu}

\usepackage{pdfpages}
\usepackage{placeins}

\makeatletter
\AtBeginDocument{\let\LS@rot\@undefined}
\makeatother

\usepackage{xr}
\usepackage{nameref}
\usepackage{varioref}
\usepackage{hyperref}
\usepackage{cleveref}

\newcommand*\cleartoleftpage{%
  \clearpage
  \ifodd\value{page}\hbox{}\newpage\fi
}

\DeclareRobustCommand{\stirling}{\genfrac\{\}{0pt}{}}

\begin{document}

\title{Bell inequalities with one bit of communication}

\author{E. Zambrini Cruzeiro and N. Gisin}

\affiliation{Department of Applied Physics, University of Geneva, 1211 Geneva, Switzerland}

\date{\today}

\begin{abstract}
We study Bell scenarios with binary outcomes supplemented by one bit of classical communication. We develop a method to find facet inequalities for such scenarios even when direct facet enumeration is not possible, or at least difficult. Using this method, we partially solve the scenario where Alice and Bob choose between three inputs, finding a total of 668 inequivalent facet inequalities (with respect to relabelings of inputs and outputs). We also show that some of these inequalities are constructed from the facet inequalities found in scenarios without communication, the well known Bell inequalities. 
\end{abstract}

\pacs{}

\maketitle

\section{Introduction}\label{sec:first}

Bell nonlocality \cite{Bell1964,Brunner2014} is one of the most intriguing phenomena encountered in modern physics. Nonlocality was discovered more than 50 years ago, and still there are simple well-posed fundamental questions about nonlocality which remain unanswered. In this article, we focus on one of these questions, which is impressively simple to state but has proven very hard to answer. In the interest of quantifying and understanding nonlocality, one can create variations of Bell's original local hidden variable (LHV) model by adding a nonlocal resource. A nonlocal resource is any resource that establishes correlations at a distance. A PR box \cite{Popescu1994,Barrett2005,Cerf2005} is an example of such a nonlocal resource. Another example is classical communication \cite{Maudlin1992,Gisin1999,Brassard1999,Steiner2000,Toner2003a}, which is the focus of this paper. In particular, one can ask how many bits of information are needed to reproduce the correlations arising from projective measurements on any two-qubit state \cite{Maudlin1992,Brassard1999,Steiner2000,Toner2003b}. For the singlet it is known that one bit is sufficient (the explicit model is given in \cite{Toner2003a}), therefore we are interested in partially entangled states, which are known to be simulable with two bits \cite{Toner2003a}, but not with zero bits \cite{Gisin1991}. We ask ourselves whether one bit also suffices to simulate projective measurements on all two-qubit partially entangled states. It is interesting that such a well-posed, binary-answer question for projective measurements on two-qubit pure states, still has not been answered even though several authors have worked on this problem \cite{Regev2009,Maxwell2014}. This illustrates the technical difficulty of studying nonlocality. Our strategy is to find Bell-like inequalities which are satisfied by all LHV models supplemented by one bit of communication, and then look for a violation of such inequalities. Although we do not provide an answer to Toner and Bacon's question here, our results already provide a deeper understanding of Bell-like inequalities for scenarios with one bit of communication.

Regular Bell scenarios and Bell scenarios supplemented with one bit of communication sent by Alice to Bob are formally described in \cref{sec:second}, along with the methods we have used to find the main results. In particular, we introduce a useful notation and propose a method to tackle scenarios where direct facet enumeration is difficult. Section \ref{sec:third} gives a proof that all projective measurements on quantum states can be reproduced by one bit of communication, for scenarios where Bob only has two dichotomic measurement settings, despite the fact that we assume the bit to be communicated from Alice to Bob. In \cref{sec:fourth}, we discuss the results we have obtained for the scenario where both Alice and Bob have three inputs. Finally, we conclude by discussing the general structure of Bell-like inequalities with one bit of communication, and future directions of research.

\section{Bell inequalities with auxiliary communication}\label{sec:second}

\subsection{Bell scenarios}\label{subsec:bell}

In a bipartite Bell scenario, the two observers are usually called Alice and Bob. Alice and Bob choose from a set of inputs (measurement settings) and as a result get an output (measurement outcome). After they have selected their inputs, Alice and Bob are not allowed to communicate. Nevertheless, they have both access to the same set of local variables because they share randomness that was generated by a common source at a past time. The observers are allowed to use the local variables to produce their outcomes. Alice and Bob both have a number of measurements settings $X, Y$ resp. and a number of outputs $A, B$. This defines the physical setup, or Bell scenario generally noted XYAB. Since in this article we restrict to binary outcome measurements, we shall note Bell scenarios XY22 simply as XY. In the lab, Alice and Bob perform measurements repetitively and record the outcome statistics, which are described by a joint probability distribution $p(ab|xy)$. If the correlations allowed by $p(ab|xy)$ are explainable using only the common past history and local operations by the observers, physicists say the experiment statistics admit a local hidden variable (LHV model). In such a case, we can write
\begin{equation}\label{eq:eq1}
p(ab|xy)=\int q(\lambda)p^A(a|x\lambda)p^B(b|y\lambda)
\end{equation}
where $\lambda$ is a local variable (infinite shared randomness), $q(\lambda)$ is its probability distribution, and $p^A(a|x\lambda),p^B(b|y\lambda)$ are respectively Alice and Bob's marginal probabilities. If Eq. \ref{eq:eq1} is not satisfied, $p(ab|xy)$ is not local.

\begin{figure}
\includegraphics[width=80mm,scale=0.3]{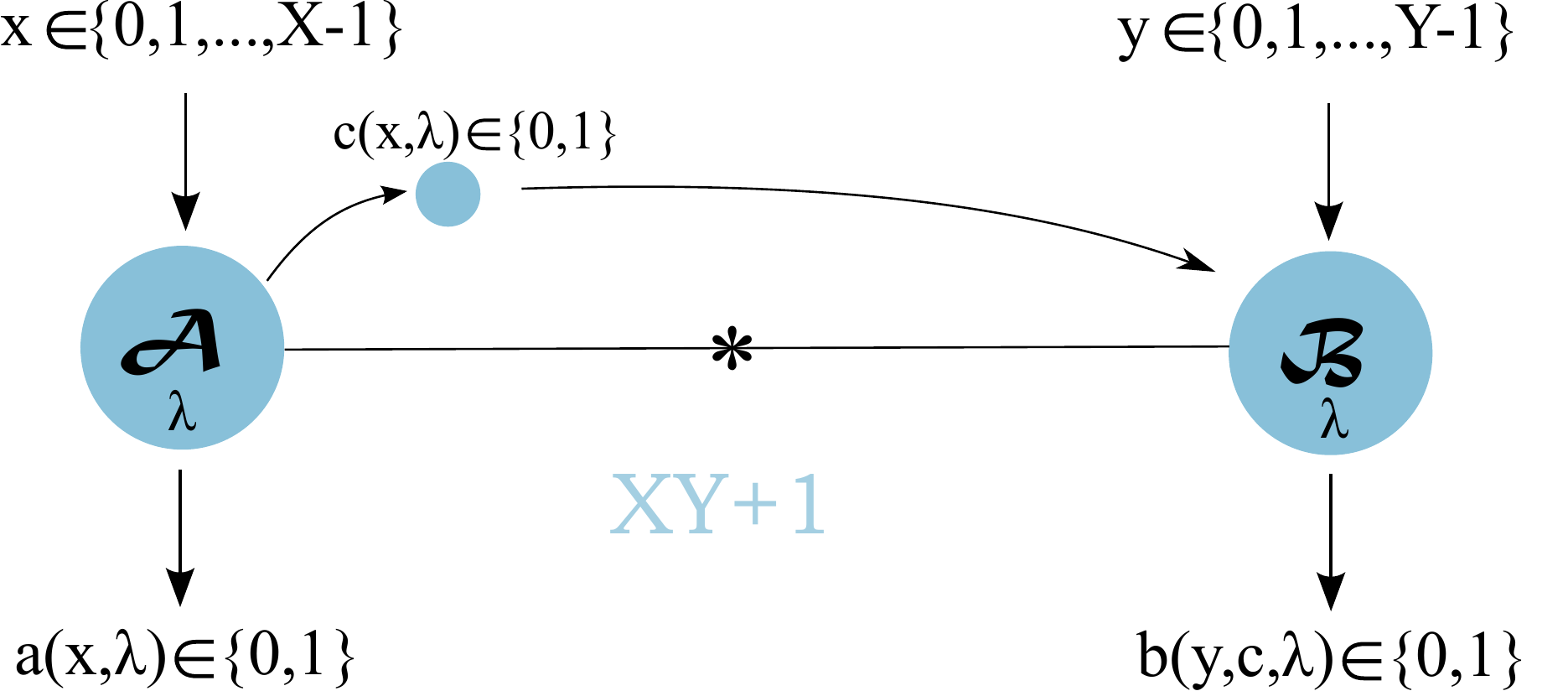}
\caption{XY+1 scenario where Alice and Bob choose between X and Y binary outcome measurements, respectively, and use share the local hidden variables $\lambda$ (shared randomness). Alice is allowed to send one bit $c(x,\lambda)$ of classical communication to Bob.}
\label{fig1}
\end{figure}

If locality is assumed, then the deterministic strategies can be defined through the marginals of Alice and Bob. The marginals define their respective local strategies. The set $\mathcal{L}$ of all local strategies $p^{\mathcal{L}}(ab|xy)$ is finite because Alice and Bob choose from a finite set of measurements, and it defines a convex polytope usually called the local polytope. The facets of this polytope define inequalities which are satisfied by any probability distribution in $\mathcal{L}$, but are violated for quantum probability distributions. These are the famous Bell inequalities, the simplest of which is the CHSH inequality, for binary inputs and outputs on both sides:
\begin{align}
\begin{split}
p(00|00)+p(00|01)+ & p(00|10)-p(00|11)\\
 & -p^A(0|0)-p^B(0|0)\leq 0
\end{split}
\end{align}
This inequality is violated by quantum mechanical probability distributions, up to $\frac{1}{\sqrt{2}}-\frac{1}{2}\approx 0.2071$.

\subsection{Bell scenarios supplemented by one cbit (Bell+1)}\label{subsec:Bellp1}

The protocol goes as follows: Alice and Bob first receive their inputs, then Alice is allowed to send one bit of classical communication to Bob. In this way Alice and Bob can simulate all $p(ab|xy)$ that satisfy: 
\begin{equation}
p(ab|xy)=\int q(\lambda)p^A(a|x\lambda)p^B(b|yc\lambda)
\end{equation}
where the marginal of Bob now depends also on the value of the classical bit $c=c(x,\lambda)$.

One can define all the local strategies with one bit of communication analogously to the original Bell scenario. The local strategies can all be written in terms of the local deterministic strategies, for which the marginal probabilities of Alice and Bob can only take the values 0 and 1. There is a finite number of such strategies, hence a finite number of vertices which define a convex polytope. Once we have generated all the vertices, we look for the facets of this polytope: this is the so-called facet enumeration problem. We will call the set of local strategies with one bit of communication $\mathcal{C}$. The inequalities defining these facets will be violated only if there exists a two-qubit state and projective measurements yielding correlations that cannot be reproduced using one bit of classical communication. 

\subsection{Local strategies for Bell+1 and notation}

The joint probability distribution $p(ab|xy)$ for each local strategy can be computed in the following way:
\begin{equation}
p(ab|xy)=\sum_{\lambda}p^A(a|x\lambda)p^B(b|cy\lambda)
\end{equation}
where $c=c(x,\lambda)$ is the communication function and can be encoded in multiple ways. In a similar fashion to Bell scenarios, we define such a scenario as XY+1, where again we omit the number of outputs as they are always binary. For a given number of inputs on Alice's side $X$, the number of communication functions in the case of one cbit is given by the Stirling number of the second kind denoted $S(X,2)$ or $\stirling{X}{2}$. The Stirling number of the second kind gives the number of distinct ways to divide a set into two non-empty subsets.

By directly generating all local strategies, we obtain $\stirling{X}{2}\cdot 2^{X}\cdot 2^{2Y}$ vertices. This method generates repeated vertices, because it takes into account the situations where Bob does not use the communication bit. By removing the repetitions we end up with a smaller number of vertices, given by
\begin{equation}\label{eq:1}
2^{X}\left(2^{Y}+\stirling{X}{2}(2^{2Y}-2^{Y})\right)
\end{equation}
This is a sum of three terms. The first term gives the vertices for the local polytope of the Bell scenario, where no communication function is used. The second term accounts for the strategies where the bit is used while the third term is for strategies where the bit is not used. An interesting consequence of this is the fact that for different values of (X,Y), one can have the same amount of vertices. In fact, any scenario XX+1 will have the same number of vertices as a (X+1)(X-1)+1 scenario. Also, any X(X+1)+1 scenario has the same number of vertices as a (X+2)(X-1)+1 scenario.

The dimension of the XY+1 polytope is X+2XY. This is the minimal number of variables (probability elements) needed to define the polytope. The usual notation for the vertices, from Toner and Bacon \cite{Toner2003b}, is given by $\{p(00|xy)\dots p(10|xy)\dots p^A(a=0|x)\dots\}$. The three dots mean that we run through all the values of x and y, for example $\{p(00|xy)\dots\}$ means $\{p(00|00),p(00|01),p(00|10),p(00|11),$ etc...$\}$. We instead chose to use the notation $\{p(00|xy)\dots p^B(b=0|xy)\dots p^A(a=0|x)\dots\}$ similarly to \cite{CollinsGisin2004} because it makes it easier to see what the inequalities reduce to when considering probability distributions in the no-signalling (NS) subspace, such as quantum probability distributions,see Table X. This becomes clear when we study the first non-trivial scenarios, 32+1 and 33+1. 2Y+1 is trivial for all Y because Alice can simply send her input as the communication bit, and in fact as we will show in section \ref{sec:third}, X2+1 is also trivial for all X.

A Bell+1 inequality can be written
\begin{equation}
\sum_{xy}d_{xy}p(00|xy)+\sum_{xy}e_{xy}p^B(0|xy)+\sum_xf_xp^A(0|x)\leq b
\end{equation}

We can represent such an inequality as a table, see \ref{tbl:table1}, in which the elements are the coefficients multiplying each of the probability elements $\{p(00|xy)\dots p^B(b=0|xy)\dots p^A(a=0|x)\dots\}$. We denote the coefficients for $p(00|xy)$ elements as $d_{xy}$, while the coefficients for Bob's marginals are $e_{xy}$ and for Alice's marginals $f_x$. Finally, an inequality is also characterized by its bound $b$.

\begin{table}[!htbp]
  \centering
  \begin{tabular}[c]{|c|ccc|}
\hline
\multirow{2}{1em}{$f_0$}&$d_{00}$&$d_{01}$&$d_{02}$\\
 &$e_{00}$&$e_{01}$&$e_{02}$\\
\hline
\multirow{2}{1em}{$f_1$}&$d_{10}$&$d_{11}$&$d_{12}$\\
 &$e_{10}$&$e_{11}$&$e_{12}$\\
\hline
\multirow{2}{1em}{$f_2$}&$d_{20}$&$d_{21}$&$d_{22}$\\
 &$e_{20}$&$e_{21}$&$e_{22}$\\
\hline
\end{tabular}$\ \leq b$
\caption{33+1 inequalities notation. $f_x$ are the weights of the marginals $p_x^A(a=0|x)$ of Alice, $d_{xy}$ are the weights of the joint probabilities for outcomes $a=b=0$ and $e_{xy}$ are the coefficients for Bob's marginals $p^B(b=0|xy)$.}%
  \label{tbl:table1}%
\end{table}

Note that a vector of this form belongs to the NS subspace if $e_{xy}$ is independent of $x$ for all $y$.

Knowing the vertices, it is possible to compute all the facets of a given polytope using dedicated software such as PORTA \cite{PORTA} or PANDA \cite{PANDA}.

\subsection{Extension of inequalities from Bell to Bell+1 scenarios and intersection of Bell+1 inequalities with the NS subspace}
\label{subsec:extension}

An inequality of a Bell scenario can be extended to the corresponding Bell+1 scenario. We extend the inequalities from the NS space to the 1-bit space by choosing the coefficients for Bob's marginals in a clever way. For any Bell inequality, there are infinitely many such extensions. We choose the one orthogonal to the NS subspace as depicted in Fig. \ref{fig3}, i.e. we impose that the vector characterising the extension lies within the NS subspace. This orthogonal extension is unique. Let us look at the example of 33+1, a scenario where we need to use this technique because a full resolution of the polytope is difficult. In table \ref{table3}, we show how to extend an arbitrary 33 inequality to the 33+1 space. We extend the inequality to the 33+1 space by adding coefficients for Bob's marginals which in this higher-dimensional space depend on both $x$ and $y$. We choose the coefficients for Bob's marginals such that $e_y'$ satisfies $e_{0y}=e_{1y}=e_{2y}=e_y'/3$ for all $y$, where $e_y'$ are the coefficients of the 33 inequality for Bob's marginals $p^B(0|y)$. In this way, one can intersect the 1-bit inequality with the non-signalling subspace and map it back to the original Bell inequality that was used for the extension. 

\begin{table}[!htbp]
\begin{tabular}{l|c c c}
       & $e_0'$ & $e_1'$ & $e_2'$ \\
      \hline
      $f_0$ & $d_{00}$ & $d_{01}$ & $d_{02}$\\
      $f_1$ & $d_{10}$ & $d_{11}$ & $d_{12}$\\
      $f_2$ & $d_{20}$ & $d_{21}$ & $d_{22}$\\
    \end{tabular}$\leq$ 0
$\ \longrightarrow \ $
\begin{tabular}[c]{|c|ccc|}
\hline
$f_0$&$d_{00}$&$d_{01}$&$d_{02}$\\
  &$e_0'/3$&$e_1'/3$&$e_2'/3$\\
  \hline
$f_1$&$d_{10}$&$d_{11}$&$d_{12}$\\
  &$e_0'/3$&$e_1'/3$&$e_2'/3$\\
  \hline
$f_2$& $d_{20}$&$d_{21}$&$d_{22}$\\
  &$e_0'/3$&$e_1'/3$&$e_2'/3$\\
\hline
\end{tabular}$\ \leq$ 0
\caption{Orthogonal extension of a Bell inequality to the 1-bit communication space (example for 33+1). The bound in both cases is the local bound.}
\label{table3}
\end{table}

Intersecting a 1-bit inequality with the NS subspace is also straightforward to do using our choice of notation, as one simply has to sum up the coefficients for Bob's marginals $\sum_{x}e_{xy}=e_y'$, then 

\begin{table}[!htbp]
  \centering
  $\vec{I}^S=$
  \begin{tabular}[c]{|c|ccc|}
\hline
\multirow{2}{1em}{$f_0$}&$d_{00}$&$d_{01}$&$d_{02}$\\
 &$e_{00}$&$e_{01}$&$e_{02}$\\
\hline
\multirow{2}{1em}{$f_1$}&$d_{10}$&$d_{11}$&$d_{12}$\\
 &$e_{10}$&$e_{11}$&$e_{12}$\\
\hline
\multirow{2}{1em}{$f_2$}&$d_{20}$&$d_{21}$&$d_{22}$\\
 &$e_{20}$&$e_{21}$&$e_{22}$\\
\hline
\end{tabular}$\ \leq b$
$\ \xrightarrow{\mathrm{NS}} \ \vec{I}^{NS}=$
\begin{tabular}{l|c c c}
       & $e_0'$ & $e_1'$ & $e_2'$ \\
      \hline
      $f_0$ & $d_{00}$ & $d_{01}$ & $d_{02}$\\
      $f_1$ & $d_{10}$ & $d_{11}$ & $d_{12}$\\
      $f_2$ & $d_{20}$ & $d_{21}$ & $d_{22}$\\
    \end{tabular}$\leq$ b
\caption{Intersecting a 1-bit inequality $I^S$ with the NS subspace amounts to summing the coefficients for Bob's marginals characterizing one of his inputs $y$.}
\label{table4}
\end{table}

\begin{figure}[!htbp]
\includegraphics[width=80mm,scale=1]{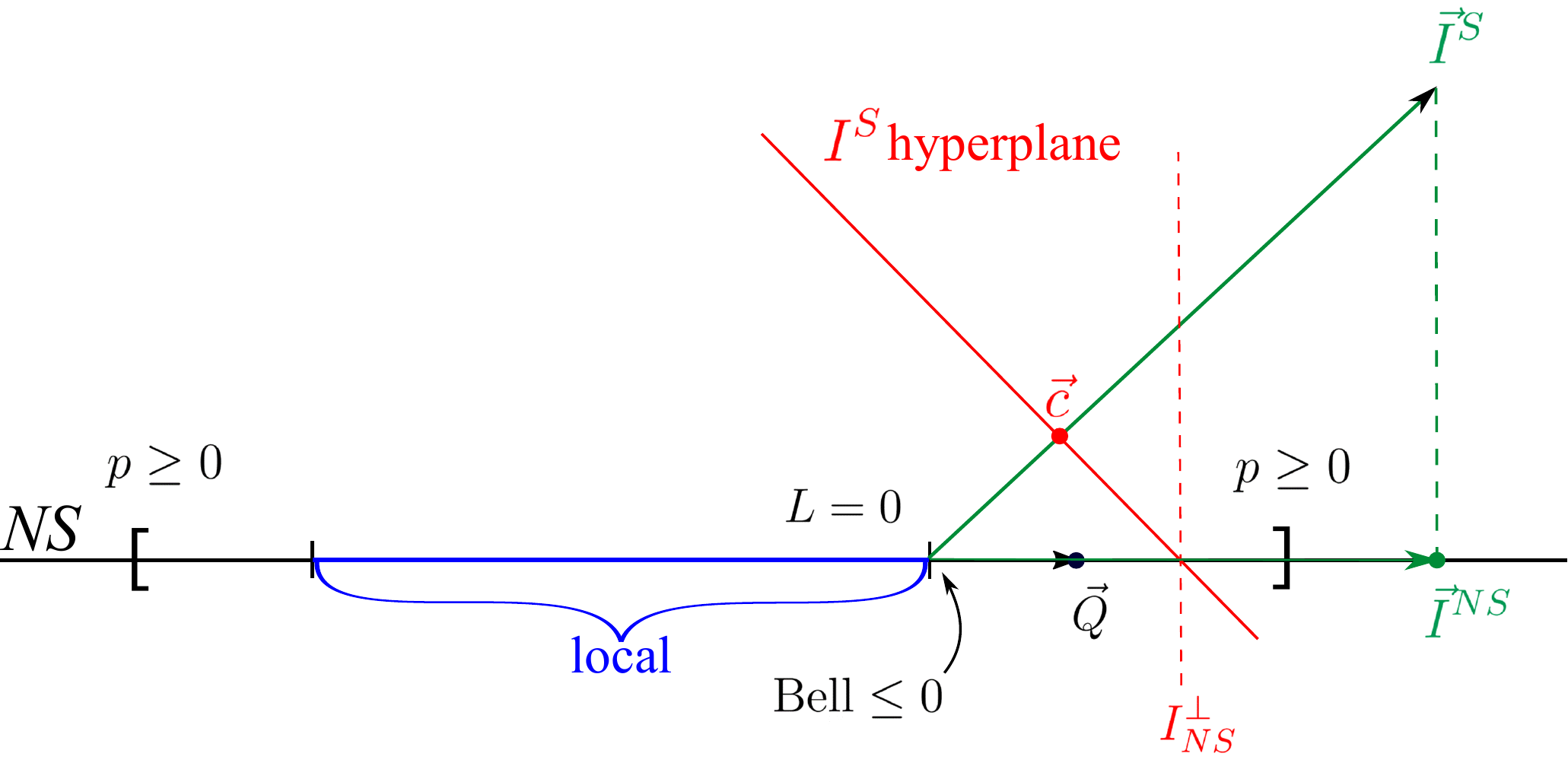}
\caption{Schematic of the geometry of the 1-bit and no-signalling spaces. The NS space is represented as a line, while the signalling space is represented as two-dimensional. The positivity conditions delimiting the NS polytope are represented by brackets. \label{fig2}}
\end{figure}

The bound for the NS inequality in Table \ref{table4} has to be considered carefully. Indeed, this bound is the 1-bit bound for $\vec{I}^S$, a particular extension (not the orthogonal one) of $\vec{I}^{NS}$ of Table \ref{table4}. Different extensions do not give the same 1-bit bound though, see Fig. \ref{fig2}. For clarity, we use a simplified scheme. In Fig. \ref{fig2}, we represent the signalling space as a plane containing the NS space, represented itself as a line. Using brackets, we also represent the bounds of the NS polytope which are given by the positivity condition $p(ab|xy)\geq 0$ for all $a$, $b$, $x$, $y$. In a similar way, the vertical lines in the NS space delimit the local polytope. The points where those lines are placed represent facets of the local polytope. A facet of the 1-bit polytope is a hyperplane $I^S$, in our representation it is an interval. In order for the probability distribution to not be reproducible by one bit of communication, we need its representative point to be further to the right than the intersection of $I^S$ with the NS space. For any point in the NS space $\vec{q}\in$ NS, $\vec{I}^S\cdot\vec{q}=\vec{I}^{NS}\cdot\vec{q}$. Therefore, a quantum bound for $\vec{I}^{NS}$ larger than the one bit bound of $\vec{I}^S$ implies that the distribution attaining the value of the quantum bound cannot be reproduced with one bit of communication. Note that the orthogonal extension's bound is always equal or larger to the correct 1-bit bound since having one bit of communication implies leaving the NS subspace.

\subsection{Cutting the polytope}
\label{subsec:cutting}

\begin{figure}[!htbp]
\includegraphics[width=80mm,scale=0.2]{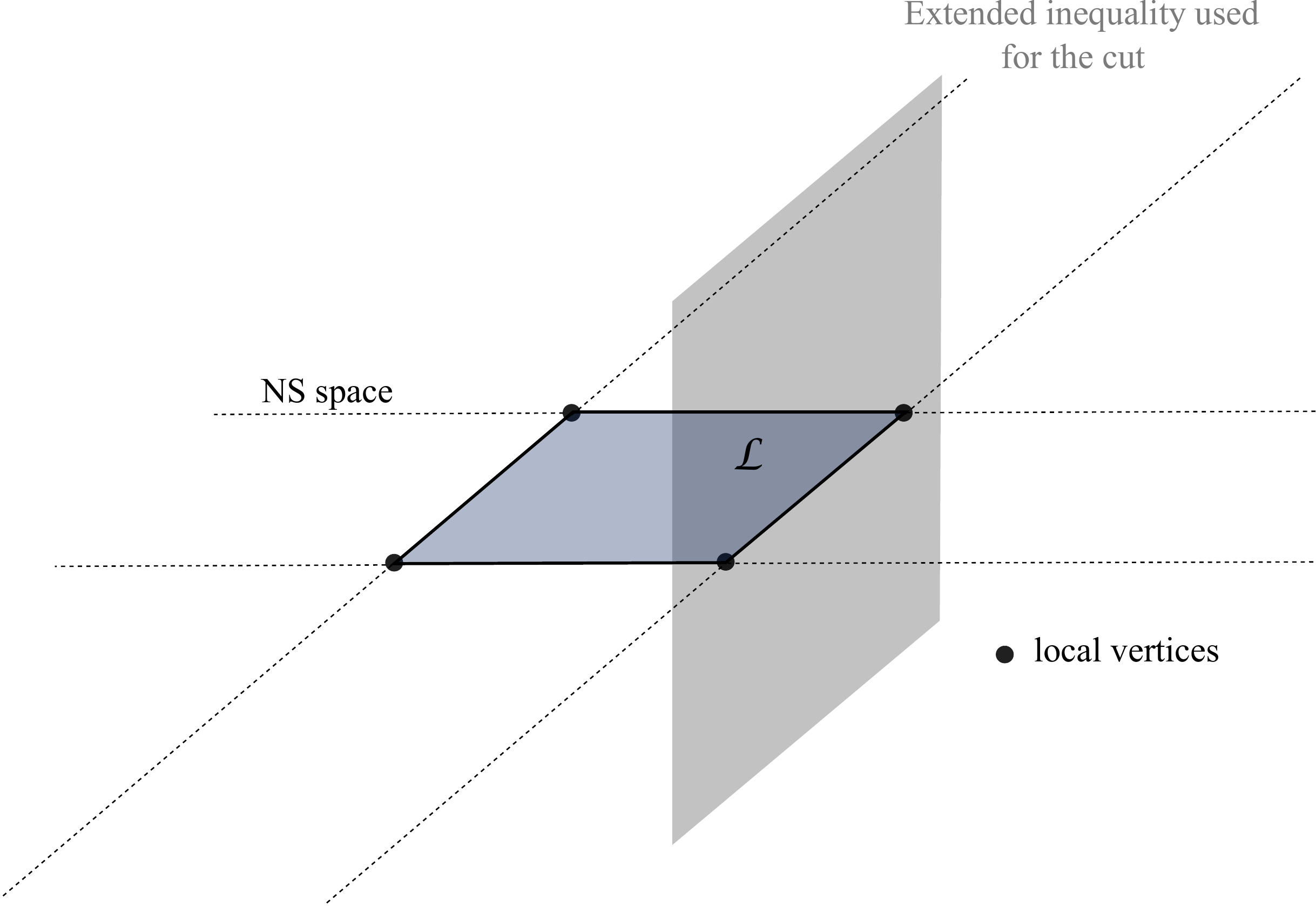}
\caption{A $\mathcal{C}$ polytope is cut by an extended Bell inequality, which is orthogonal to the NS subspace. The NS subspace is represented as a two-dimensional space. We choose not to represent the $\mathcal{C}$ polytope as we do not know it's geometrical form. By keeping all the vertices which saturate or violate such an inequality one obtains a subpolytope for which it is easier to find the facets via direct facet enumeration.\label{fig3}}
\end{figure}

When direct facet enumeration cannot be done in up to one or two weeks, we use a trick to find a smaller set of inequalities. The trick consists in enumerating the facets for a subpolytope of $\mathcal{C}$, where $\mathcal{C}$ is the 1-bit polytope. The way we select the subpolytope is by taking a Bell scenario inequality, extending it to the 1-bit space in an orthogonal way as shown in Fig.~\ref{fig3} and removing any vertex that satisfies this new inequality. This amounts to cutting the polytope with a hyperplane.

As previously described, we choose the coefficients for Bob's marginals in the 1-bit space to be equal because this corresponds to an orthogonal extension of the facet with respect to the NS space, i.e. $I^{\mathrm{S}}\perp\mathrm{NS}$, where $I^{\mathrm{S}}$ is the rightmost inequality in Table \ref{table3}. We have tested the choice of the coefficients extensively with the 32+1 scenario, which has already been fully solved \cite{Maxwell2014}. In order to generate all the relevant facets, it is important that Bob's marginals for the inputs which give a CHSH inequality are equal. The other coefficients seem completely arbitrary. In the 33+1 example of table \ref{table3}, for Bob's input $y=1$, this means the coefficients for $p^B(0|xy)$ for $x=0,1$ should be equal, and the coefficient for $x=2$ is  arbitrary. 

When we change the choice of coefficients for Bob's marginals, we are performing a rotation of the hyperplane used to cut the 1-bit polytope. Therefore, one could try different choices of coefficients in order to select different sets of vertices and therefore produce several subpolytopes out of the original polytope. Furthermore, each relabelling of the inequality cuts a different region of the polytope, possibly revealing new facets.

There is another freedom for the cut, one can modify the bound of the inequality used for the cut. This will cause a translation of the hyperplane, which allows one to change the size of the subpolytopes we generate. Therefore for very hard problems we can increase the bound to try to solve smaller subpolytopes. This translation technique has been used before, see \cite{Pal2009} for further details. 

Last but not least, when we cut a polytope and find the facets of the subpolytope, some facets are not facets of the original polytope: they were created by the cut. In order to keep only the relevant inequalities, we check their rank and whether the vertices of the original polytope saturate exactly the bound of the inequalities.

\section{X2+1 scenarios}
\label{sec:third}

2Y+1 scenarios are trivial because Alice can send her input as the classical bit. Surprisingly, X2+1 inequalities also cannot be violated by any NS distribution despite the fact that we assume the classical bit is sent from Alice to Bob. The reason is that every NS vertex of a X2 Bell scenario can be reproduced using one PR box \cite{Jones2005,Barrett2005}. Therefore, one PR box can simulate any quantum state in X2 scenarios, as boxes can be written as convex combinations of the NS vertices. Furthermore, one bit of communication is a strictly stronger nonlocal resource that one PR box \cite{Gisin2013}. Therefore one bit of communication can simulate any quantum state in a X2 scenario.

\section{33+1 scenario}
\label{sec:fourth}

In this section, we present our results for the 33+1 scenario. For this scenario the facet enumeration is demanding, but by cutting the polytope one is able to recover a large list of inequalities. In the corresponding 33 Bell scenario, besides CHSH there is one new inequality, called $I_{\mathrm{3322}}$, which we can also use to cut the 33+1 polytope:

\begin{table}[!htbp]
    $I_{3322}=$\begin{tabular}{l|c c c} 
       & -1 & 0 & 0 \\
      \hline
      -2 & 1 & 1 & 1\\
      -1 & 1 & 1 & -1\\
      0  & 1 & -1 & 0
    \end{tabular}$\leq$ 0
\label{tableXuts}
\end{table}


\subsection{Cutting with CHSH}

We apply the cut with the extended CHSH inequality, using the procedure described above. 

We then solve the subpolytope. We find 657 inequivalent inequalities, where 179 inequalities have a quantum advantage when intersected with the NS subspace. Note that quantum probabilities do not violate the 1-bit bound $C$, but they can offer, as is the case for the 179 inequalities, an advantage with respect to the local bound $L$ in the NS subspace. One can distinguish the inequalities by how close the quantum bound is from the one bit bound, with the following figure of merit:
\begin{equation}
\frac{Q-L}{C-L}
\end{equation}
This figure of merit also gives a lower bound on the amount of average communication required to reproduce 3322 correlations \cite{Pironio2003}. The best quantum bound that we obtain with respect to the local bound is halfway between the local and 1-bit bounds, see inequalities 195 and 232 in Table X. This result implies that to reproduce 3322 correlations, Alice needs to send to Bob one bit at least half of the time on average. By looking in the NS subspace, we can show that our half-way quantum bound is obtained through a sum of two $I_{3322}$ inequalities (recall that the quantum bound of $I_{3322}$ is equal to 0.25). We also find inequalities which in the NS subspace correspond to the sum of two CHSH, and inequalities corresponding to one CHSH or one $I_{3322}$. We also find violations which correspond to a CHSH or an $I_{3322}$ inequality plus some term which changes the optimal state/measurement and therefore modifies the quantum bound too. Doing the same analysis in the 32+1 scenario, one finds that correlations can be reproduced only if the amount of average communication is higher than 0.4142.

\begin{table}[!htbp]
\begin{tabular}[c]{|c|ccc|}
\hline
-3&2&2&2\\
  &-1&-1&-1\\
  \hline
-1&2&2&-2\\
  &-1&-1&1\\
  \hline
0& 2&-2&0\\
  &-1&1&0\\
\hline
\end{tabular}$\ \leq$ 1
$\ \xrightarrow{\mathrm{NS}} \ $
\begin{tabular}{l|c c c}
       & -3 & -1 & 0 \\
      \hline
      -3 & 2 & 2 & 2\\
      -1 & 2 & 2 & -2\\
      0 & 2 & -2 & 0\\
    \end{tabular}$\leq$ 1
\caption{Facet of 33+1, for which the quantum bound is halfway between the local and 1-bit bounds. When intersected with the NS space, this inequality reduces to a sum of $I_{3322}$ inequalities. This inequality corresponds to facet number 232 in Table X.}
\label{table5}
\end{table}

In table \ref{table5}, we give an explicit example of a facet which has a larger quantum bound with respect to the local bound: inequality number 232 in Table X, which can be found in the Appendix. In the non-signalling subspace, this facet corresponds to a sum of $I_{3322}$. In order to make this clear, we intersect the facet of the $\mathcal{C}$ polytope with the NS subspace.


The resulting inequality is $I_{3322}+I_{3322}^{\mathrm{perm}}$ with a bound of one instead of zero, where $I_{3322}^{\mathrm{perm}}$ is $I_{3322}$ with a relabeling of the parties (permutation of Alice and Bob labels). We found another inequality of the same type, which also includes a sum of $I_{3322}$ and $I_{3322}^{\mathrm{perm}}$, although it is less obvious to see it because it also includes some other terms which do not contribute to the quantum bound. The second inequality (number 195 in Table X) is

\begin{table}[!htbp]
\begin{tabular}[c]{|c|ccc|}
\hline
-3&2&2&2\\
  &-1&-1&-1\\
  \hline
-1&2&-2&1\\
  &-1&1&0\\
  \hline
0& 2&1&-2\\
  &-1&-1&1\\
\hline
\end{tabular}$\ \leq$ 1
$\ \xrightarrow{\mathrm{NS}} \ $
\begin{tabular}{l|c c c}
       & -3 & -1 & 0 \\
      \hline
      -3 & 2 & 2 & 2\\
      -1 & 2 & -2 & 1\\
      0 & 2 & 1 & -2\\
    \end{tabular}$\leq$ 1
    \caption{Second facet (number 195) of 33+1 for which the quantum bound is halfway between the local and 1-bit bounds.}
\label{table6}
\end{table}

We give more examples of 33+1 inequalities in the Appendix, along with their NS intersections.

We have also tested the subpolytope method in the $32+1$ scenario. By cutting with CHSH, we retrieve 80 inequalities. By removing those which are not true facets of the 1-bit polytope, we obtain 17 inequalities. By sorting these inequalities into inequivalence classes, we end up with 9 inequalities, a positivity facet and the 8 new facets which were published in \cite{Maxwell2014}. In this scenario, by cutting the polytope one easily recovers the complete list of facet inequalities. Additionally, by intersecting these facets with the NS subspace, we find again that the inequalities which have a larger quantum bound than local bound are constructed from CHSH. The best inequality in terms of distance between the local and quantum bounds in 32+1 is a sum of two CHSH.

\subsection{Cutting with I3322}

We repeat the ``cutting'' procedure, using the $I_{3322}$ inequality instead of CHSH. There are two other versions (in fact many more: any relabeling as discussed in Section \ref{subsec:cutting}) of $I_{3322}$ that we can use. One of them is $I_{3322}^{\mathrm{perm}}$, which we have introduced previously. The other is the symmetrized version of $I_{3322}$:

\begin{table}[!htbp]
    $I_{3322}^{\mathrm{sym}}=$\begin{tabular}{l|c c c} 
       & -1 & -1 & 0 \\
      \hline
      -1 & 0 & 1 & 1\\
      -1 & 1 & -1 & 1\\
      0  & 1 & 1 & -1
    \end{tabular}$\leq$ 0
\label{table7}
\end{table}

These inequalities are equivalent in the NS subspace, but when extended to the 1-bit space they become inequivalent. Therefore each cut will give a different number of vertices and facets. Cutting with $I_{3322}$, we obtain 513 inequivalent facets, 151 of them having a larger quantum bound than local bound in the NS subspace. The cut with $I_{3322}^{\mathrm{sym}}$ yields 642 inequivalent inequalities, 171 of them having a quantum advantage in the NS subspace. Finally $I_{3322}^{\mathrm{perm}}$ gives 634 facets, 174 with a quantum advantage in the NS subspace.

We grouped all these inequalities together, and removed equivalent inequalities. We end up with a total of 667 inequalities, 184 of which have a stronger quantum bound than local bound.

We find the same construction as before, the inequalities are constructed out of inequalities of the Bell polytope. For example, we find the same facet inequalities for $33+1$ which reduce to the sum of two $I_{3322}$ in the NS subspace.

We have also attempted to directly solve the full polytope. At the moment when we extracted the list of inequalities generated with the full polytope, the number of inequalities had not increased in the last two months. We thus conjecture that the list of 668 facet inequalities is complete.

\section{Conclusion}

We present a method and notation to find facets of Bell scenarios supplemented by one bit of classical communication. The notation we use simplifies the study of the 1-bit inequalities, especially with respect to their intersection with the NS subspace. Even though the 1-bit polytope is difficult to solve directly, we are able to find an extensive list of facets, that we conjecture to be complete. In the scenario $33+1$, we find no quantum violation of the 1-bit bound. Given the structure of $33+1$ facets, and assuming our conjecture is correct, we have proven that the statistics obtained by choosing between three projective measurements on any two-qubit quantum state can be reproduced by one bit of classical communication between the parties. Our results also imply that in this scenario, Alice must send one bit at least half of the time on average to Bob in order for the two parties to reproduce the quantum correlations. These findings constitute a step further towards answering the binary-answer question raised in Section \ref{sec:first}. Our results provide a better understanding of the general structure of Bell inequalities supplemented by one bit. Indeed, we found that by intersecting the facets of the $\mathcal{C}$ polytope with the NS subspace, we derive inequalities which are constructed from Bell inequalities of the corresponding scenario without the communication. This can be a starting point to guess new facets for scenarios where the Bell inequalities are known.

The next scenarios to tackle are 34+1, 43+1 and 44+1. An important point is that our results show the best inequalities we find in terms of distance between the local and quantum bounds are sums of the same Bell inequality of the corresponding Bell scenario, for example for 33+1 the best inequality is a sum of two Bell inequalities from 33. If this is a general trend for Bell scenarios supplemented by one bit of communication, then in order to find a violation of the 1-bit bound we require that the Bell inequalities of the corresponding Bell scenario should be:
\begin{itemize}
 \item[1)] maximally violated by a partially entangled state.\\
 \item[2)] have a quantum bound which is more than half-way between the local and 1-bit bounds.
\end{itemize} 

Only starting from four settings on one side and three on the other do we have partially entangled states maximally violating a Bell inequality \cite{Brunner2008}. In addition, in the 44+1 scenario the states which maximally violate the Bell inequalities are in most of the cases very close to maximally entangled \cite{Brunner2008}. Furthermore, for polytopes of higher dimension than the 44 scenario \cite{ZambriniCruzeiro2018}, we still do not know the complete list of facets, which makes the problem even more complicated. All of these points are quite negative in the perspective of solving the binary-answer question, nevertheless there are possible avenues to get closer to the solution. One idea is to generate facets from subpolytopes of such complicated scenarios, but one has to be lucky to find the optimal inequalities in terms of communication. Another possibility is to guess inequalities using known Bell inequalities, at least up to four settings for each party.

\begin{acknowledgments}
The authors would like to thank S. Pironio, F. Hirsch, D. Rosset and N. Brunner for useful discussions. Financial support by the Swiss NCCR-QSIT is gratefully
acknowledged.
\end{acknowledgments}

\bibliography{1cbit}

\begin{thebibliography}{23}%
\makeatletter
\providecommand \@ifxundefined [1]{%
 \@ifx{#1\undefined}
}%
\providecommand \@ifnum [1]{%
 \ifnum #1\expandafter \@firstoftwo
 \else \expandafter \@secondoftwo
 \fi
}%
\providecommand \@ifx [1]{%
 \ifx #1\expandafter \@firstoftwo
 \else \expandafter \@secondoftwo
 \fi
}%
\providecommand \natexlab [1]{#1}%
\providecommand \enquote  [1]{``#1''}%
\providecommand \bibnamefont  [1]{#1}%
\providecommand \bibfnamefont [1]{#1}%
\providecommand \citenamefont [1]{#1}%
\providecommand \href@noop [0]{\@secondoftwo}%
\providecommand \href [0]{\begingroup \@sanitize@url \@href}%
\providecommand \@href[1]{\@@startlink{#1}\@@href}%
\providecommand \@@href[1]{\endgroup#1\@@endlink}%
\providecommand \@sanitize@url [0]{\catcode `\\12\catcode `\$12\catcode
  `\&12\catcode `\#12\catcode `\^12\catcode `\_12\catcode `\%12\relax}%
\providecommand \@@startlink[1]{}%
\providecommand \@@endlink[0]{}%
\providecommand \url  [0]{\begingroup\@sanitize@url \@url }%
\providecommand \@url [1]{\endgroup\@href {#1}{\urlprefix }}%
\providecommand \urlprefix  [0]{URL }%
\providecommand \Eprint [0]{\href }%
\providecommand \doibase [0]{http://dx.doi.org/}%
\providecommand \selectlanguage [0]{\@gobble}%
\providecommand \bibinfo  [0]{\@secondoftwo}%
\providecommand \bibfield  [0]{\@secondoftwo}%
\providecommand \translation [1]{[#1]}%
\providecommand \BibitemOpen [0]{}%
\providecommand \bibitemStop [0]{}%
\providecommand \bibitemNoStop [0]{.\EOS\space}%
\providecommand \EOS [0]{\spacefactor3000\relax}%
\providecommand \BibitemShut  [1]{\csname bibitem#1\endcsname}%
\let\auto@bib@innerbib\@empty
\bibitem [{\citenamefont {Bell}(1964)}]{Bell1964}%
  \BibitemOpen
  \bibfield  {author} {\bibinfo {author} {\bibfnamefont {J.~S.}\ \bibnamefont
  {Bell}},\ }\href@noop {} {\bibfield  {journal} {\bibinfo  {journal} {Google
  Scholar}\ } (\bibinfo {year} {1964})}\BibitemShut {NoStop}%
\bibitem [{\citenamefont {Brunner}\ \emph {et~al.}(2014)\citenamefont
  {Brunner}, \citenamefont {Cavalcanti}, \citenamefont {Pironio}, \citenamefont
  {Scarani},\ and\ \citenamefont {Wehner}}]{Brunner2014}%
  \BibitemOpen
  \bibfield  {author} {\bibinfo {author} {\bibfnamefont {N.}~\bibnamefont
  {Brunner}}, \bibinfo {author} {\bibfnamefont {D.}~\bibnamefont {Cavalcanti}},
  \bibinfo {author} {\bibfnamefont {S.}~\bibnamefont {Pironio}}, \bibinfo
  {author} {\bibfnamefont {V.}~\bibnamefont {Scarani}}, \ and\ \bibinfo
  {author} {\bibfnamefont {S.}~\bibnamefont {Wehner}},\ }\href@noop {}
  {\bibfield  {journal} {\bibinfo  {journal} {Reviews of Modern Physics}\
  }\textbf {\bibinfo {volume} {86}},\ \bibinfo {pages} {419} (\bibinfo {year}
  {2014})}\BibitemShut {NoStop}%
\bibitem [{\citenamefont {Popescu}\ and\ \citenamefont
  {Rohrlich}(1994)}]{Popescu1994}%
  \BibitemOpen
  \bibfield  {author} {\bibinfo {author} {\bibfnamefont {S.}~\bibnamefont
  {Popescu}}\ and\ \bibinfo {author} {\bibfnamefont {D.}~\bibnamefont
  {Rohrlich}},\ }\href@noop {} {\bibfield  {journal} {\bibinfo  {journal}
  {Foundations of Physics}\ }\textbf {\bibinfo {volume} {24}},\ \bibinfo
  {pages} {379} (\bibinfo {year} {1994})}\BibitemShut {NoStop}%
\bibitem [{\citenamefont {Barrett}\ and\ \citenamefont
  {Pironio}(2005)}]{Barrett2005}%
  \BibitemOpen
  \bibfield  {author} {\bibinfo {author} {\bibfnamefont {J.}~\bibnamefont
  {Barrett}}\ and\ \bibinfo {author} {\bibfnamefont {S.}~\bibnamefont
  {Pironio}},\ }\href@noop {} {\bibfield  {journal} {\bibinfo  {journal}
  {Physical review letters}\ }\textbf {\bibinfo {volume} {95}},\ \bibinfo
  {pages} {140401} (\bibinfo {year} {2005})}\BibitemShut {NoStop}%
\bibitem [{\citenamefont {Cerf}\ \emph {et~al.}(2005)\citenamefont {Cerf},
  \citenamefont {Gisin}, \citenamefont {Massar},\ and\ \citenamefont
  {Popescu}}]{Cerf2005}%
  \BibitemOpen
  \bibfield  {author} {\bibinfo {author} {\bibfnamefont {N.~J.}\ \bibnamefont
  {Cerf}}, \bibinfo {author} {\bibfnamefont {N.}~\bibnamefont {Gisin}},
  \bibinfo {author} {\bibfnamefont {S.}~\bibnamefont {Massar}}, \ and\ \bibinfo
  {author} {\bibfnamefont {S.}~\bibnamefont {Popescu}},\ }\href@noop {}
  {\bibfield  {journal} {\bibinfo  {journal} {Physical Review Letters}\
  }\textbf {\bibinfo {volume} {94}},\ \bibinfo {pages} {220403} (\bibinfo
  {year} {2005})}\BibitemShut {NoStop}%
\bibitem [{\citenamefont {Maudlin}(1992)}]{Maudlin1992}%
  \BibitemOpen
  \bibfield  {author} {\bibinfo {author} {\bibfnamefont {T.}~\bibnamefont
  {Maudlin}},\ }in\ \href@noop {} {\emph {\bibinfo {booktitle} {PSA:
  Proceedings of the Biennial Meeting of the Philosophy of Science
  Association}}},\ Vol.\ \bibinfo {volume} {1992}\ (\bibinfo {organization}
  {Philosophy of Science Association},\ \bibinfo {year} {1992})\ pp.\ \bibinfo
  {pages} {404--417}\BibitemShut {NoStop}%
\bibitem [{\citenamefont {Gisin}\ and\ \citenamefont
  {Gisin}(1999)}]{Gisin1999}%
  \BibitemOpen
  \bibfield  {author} {\bibinfo {author} {\bibfnamefont {N.}~\bibnamefont
  {Gisin}}\ and\ \bibinfo {author} {\bibfnamefont {B.}~\bibnamefont {Gisin}},\
  }\href@noop {} {\bibfield  {journal} {\bibinfo  {journal} {Physics Letters
  A}\ }\textbf {\bibinfo {volume} {260}},\ \bibinfo {pages} {323} (\bibinfo
  {year} {1999})}\BibitemShut {NoStop}%
\bibitem [{\citenamefont {Brassard}\ \emph {et~al.}(1999)\citenamefont
  {Brassard}, \citenamefont {Cleve},\ and\ \citenamefont
  {Tapp}}]{Brassard1999}%
  \BibitemOpen
  \bibfield  {author} {\bibinfo {author} {\bibfnamefont {G.}~\bibnamefont
  {Brassard}}, \bibinfo {author} {\bibfnamefont {R.}~\bibnamefont {Cleve}}, \
  and\ \bibinfo {author} {\bibfnamefont {A.}~\bibnamefont {Tapp}},\ }\href@noop
  {} {\bibfield  {journal} {\bibinfo  {journal} {Physical Review Letters}\
  }\textbf {\bibinfo {volume} {83}},\ \bibinfo {pages} {1874} (\bibinfo {year}
  {1999})}\BibitemShut {NoStop}%
\bibitem [{\citenamefont {Steiner}(2000)}]{Steiner2000}%
  \BibitemOpen
  \bibfield  {author} {\bibinfo {author} {\bibfnamefont {M.}~\bibnamefont
  {Steiner}},\ }\href@noop {} {\bibfield  {journal} {\bibinfo  {journal}
  {Physics Letters A}\ }\textbf {\bibinfo {volume} {270}},\ \bibinfo {pages}
  {239} (\bibinfo {year} {2000})}\BibitemShut {NoStop}%
\bibitem [{\citenamefont {Toner}\ and\ \citenamefont
  {Bacon}(2003)}]{Toner2003a}%
  \BibitemOpen
  \bibfield  {author} {\bibinfo {author} {\bibfnamefont {B.~F.}\ \bibnamefont
  {Toner}}\ and\ \bibinfo {author} {\bibfnamefont {D.}~\bibnamefont {Bacon}},\
  }\href@noop {} {\bibfield  {journal} {\bibinfo  {journal} {Physical Review
  Letters}\ }\textbf {\bibinfo {volume} {91}},\ \bibinfo {pages} {187904}
  (\bibinfo {year} {2003})}\BibitemShut {NoStop}%
\bibitem [{\citenamefont {Bacon}\ and\ \citenamefont
  {Toner}(2003)}]{Toner2003b}%
  \BibitemOpen
  \bibfield  {author} {\bibinfo {author} {\bibfnamefont {D.}~\bibnamefont
  {Bacon}}\ and\ \bibinfo {author} {\bibfnamefont {B.~F.}\ \bibnamefont
  {Toner}},\ }\href@noop {} {\bibfield  {journal} {\bibinfo  {journal}
  {Physical review letters}\ }\textbf {\bibinfo {volume} {90}},\ \bibinfo
  {pages} {157904} (\bibinfo {year} {2003})}\BibitemShut {NoStop}%
\bibitem [{\citenamefont {Gisin}(1991)}]{Gisin1991}%
  \BibitemOpen
  \bibfield  {author} {\bibinfo {author} {\bibfnamefont {N.}~\bibnamefont
  {Gisin}},\ }\href@noop {} {\bibfield  {journal} {\bibinfo  {journal} {Physics
  Letters A}\ }\textbf {\bibinfo {volume} {154}},\ \bibinfo {pages} {201}
  (\bibinfo {year} {1991})}\BibitemShut {NoStop}%
\bibitem [{\citenamefont {Regev}\ and\ \citenamefont
  {Toner}(2009)}]{Regev2009}%
  \BibitemOpen
  \bibfield  {author} {\bibinfo {author} {\bibfnamefont {O.}~\bibnamefont
  {Regev}}\ and\ \bibinfo {author} {\bibfnamefont {B.}~\bibnamefont {Toner}},\
  }\href@noop {} {\bibfield  {journal} {\bibinfo  {journal} {SIAM Journal on
  Computing}\ }\textbf {\bibinfo {volume} {39}},\ \bibinfo {pages} {1562}
  (\bibinfo {year} {2009})}\BibitemShut {NoStop}%
\bibitem [{\citenamefont {Maxwell}\ and\ \citenamefont
  {Chitambar}(2014)}]{Maxwell2014}%
  \BibitemOpen
  \bibfield  {author} {\bibinfo {author} {\bibfnamefont {K.}~\bibnamefont
  {Maxwell}}\ and\ \bibinfo {author} {\bibfnamefont {E.}~\bibnamefont
  {Chitambar}},\ }\href@noop {} {\bibfield  {journal} {\bibinfo  {journal}
  {Physical Review A}\ }\textbf {\bibinfo {volume} {89}},\ \bibinfo {pages}
  {042108} (\bibinfo {year} {2014})}\BibitemShut {NoStop}%
\bibitem [{\citenamefont {Collins}\ and\ \citenamefont
  {Gisin}(2004)}]{CollinsGisin2004}%
  \BibitemOpen
  \bibfield  {author} {\bibinfo {author} {\bibfnamefont {D.}~\bibnamefont
  {Collins}}\ and\ \bibinfo {author} {\bibfnamefont {N.}~\bibnamefont
  {Gisin}},\ }\href@noop {} {\bibfield  {journal} {\bibinfo  {journal} {Journal
  of Physics A: Mathematical and General}\ }\textbf {\bibinfo {volume} {37}},\
  \bibinfo {pages} {1775} (\bibinfo {year} {2004})}\BibitemShut {NoStop}%
\bibitem [{POR()}]{PORTA}%
  \BibitemOpen
  \href {https://wwwproxy.iwr.uni-heidelberg.de/groups/comopt/software/PORTA/}
  {\enquote {\bibinfo {title}
  {https://wwwproxy.iwr.uni-heidelberg.de/groups/comopt/software/porta/},}\
  }\BibitemShut {NoStop}%
\bibitem [{\citenamefont {Lörwald}\ and\ \citenamefont
  {Reinelt}(2015)}]{PANDA}%
  \BibitemOpen
  \bibfield  {author} {\bibinfo {author} {\bibfnamefont {S.}~\bibnamefont
  {Lörwald}}\ and\ \bibinfo {author} {\bibfnamefont {G.}~\bibnamefont
  {Reinelt}},\ }\href {\doibase 10.1007/s13675-015-0040-0} {\bibfield
  {journal} {\bibinfo  {journal} {EURO Journal on Computational Optimization}\
  ,\ \bibinfo {pages} {1}} (\bibinfo {year} {2015})}\BibitemShut {NoStop}%
\bibitem [{\citenamefont {P{\'a}l}\ and\ \citenamefont
  {V{\'e}rtesi}(2009)}]{Pal2009}%
  \BibitemOpen
  \bibfield  {author} {\bibinfo {author} {\bibfnamefont {K.~F.}\ \bibnamefont
  {P{\'a}l}}\ and\ \bibinfo {author} {\bibfnamefont {T.}~\bibnamefont
  {V{\'e}rtesi}},\ }\href@noop {} {\bibfield  {journal} {\bibinfo  {journal}
  {Physical Review A}\ }\textbf {\bibinfo {volume} {79}},\ \bibinfo {pages}
  {022120} (\bibinfo {year} {2009})}\BibitemShut {NoStop}%
\bibitem [{\citenamefont {Jones}\ and\ \citenamefont
  {Masanes}(2005)}]{Jones2005}%
  \BibitemOpen
  \bibfield  {author} {\bibinfo {author} {\bibfnamefont {N.~S.}\ \bibnamefont
  {Jones}}\ and\ \bibinfo {author} {\bibfnamefont {L.}~\bibnamefont
  {Masanes}},\ }\href@noop {} {\bibfield  {journal} {\bibinfo  {journal}
  {Physical Review A}\ }\textbf {\bibinfo {volume} {72}},\ \bibinfo {pages}
  {052312} (\bibinfo {year} {2005})}\BibitemShut {NoStop}%
\bibitem [{\citenamefont {Gisin}\ \emph {et~al.}(2013)\citenamefont {Gisin},
  \citenamefont {Popescu}, \citenamefont {Scarani}, \citenamefont {Wolf},\ and\
  \citenamefont {Wullschleger}}]{Gisin2013}%
  \BibitemOpen
  \bibfield  {author} {\bibinfo {author} {\bibfnamefont {N.}~\bibnamefont
  {Gisin}}, \bibinfo {author} {\bibfnamefont {S.}~\bibnamefont {Popescu}},
  \bibinfo {author} {\bibfnamefont {V.}~\bibnamefont {Scarani}}, \bibinfo
  {author} {\bibfnamefont {S.}~\bibnamefont {Wolf}}, \ and\ \bibinfo {author}
  {\bibfnamefont {J.}~\bibnamefont {Wullschleger}},\ }\href@noop {} {\bibfield
  {journal} {\bibinfo  {journal} {Natural Computing}\ }\textbf {\bibinfo
  {volume} {12}},\ \bibinfo {pages} {13} (\bibinfo {year} {2013})}\BibitemShut
  {NoStop}%
\bibitem [{\citenamefont {Pironio}(2003)}]{Pironio2003}%
  \BibitemOpen
  \bibfield  {author} {\bibinfo {author} {\bibfnamefont {S.}~\bibnamefont
  {Pironio}},\ }\href@noop {} {\bibfield  {journal} {\bibinfo  {journal}
  {Physical Review A}\ }\textbf {\bibinfo {volume} {68}},\ \bibinfo {pages}
  {062102} (\bibinfo {year} {2003})}\BibitemShut {NoStop}%
\bibitem [{\citenamefont {Brunner}\ and\ \citenamefont
  {Gisin}(2008)}]{Brunner2008}%
  \BibitemOpen
  \bibfield  {author} {\bibinfo {author} {\bibfnamefont {N.}~\bibnamefont
  {Brunner}}\ and\ \bibinfo {author} {\bibfnamefont {N.}~\bibnamefont
  {Gisin}},\ }\href@noop {} {\bibfield  {journal} {\bibinfo  {journal} {Physics
  Letters A}\ }\textbf {\bibinfo {volume} {372}},\ \bibinfo {pages} {3162}
  (\bibinfo {year} {2008})}\BibitemShut {NoStop}%
\bibitem [{\citenamefont {Cruzeiro}\ and\ \citenamefont
  {Gisin}(2018)}]{ZambriniCruzeiro2018}%
  \BibitemOpen
  \bibfield  {author} {\bibinfo {author} {\bibfnamefont {E.~Z.}\ \bibnamefont
  {Cruzeiro}}\ and\ \bibinfo {author} {\bibfnamefont {N.}~\bibnamefont
  {Gisin}},\ }\href@noop {} {\bibfield  {journal} {\bibinfo  {journal} {arXiv
  preprint arXiv:1811.11820}\ } (\bibinfo {year} {2018})}\BibitemShut {NoStop}%
\end{thebibliography}%

\clearpage

\appendix

\section*{Examples of 33+1 facets and complete list}
\label{sec:appendix}

In this appendix, we give examples of facets of 33+1, along with their NS intersection and their connection to Bell inequalities. We start with a 33+1 facet which reduces to the sum of two CHSH inequalities in the NS subspace, inequality 349 in our Table X.

\begin{table}[!htbp]
\begin{tabular}[c]{|c|ccc|}
\hline
0&-1&0&1\\
  &0&0&0\\
  \hline
0&0&1&-1\\
  &0&0&0\\
  \hline
-2& 1&1&2\\
  &0&-1&-1\\
\hline
\end{tabular}$\ \leq$ 1
$\ \xrightarrow{\mathrm{NS}} \ $
\begin{tabular}{l|c c c}
       & 0 & -1 & -1 \\
      \hline
      0 & -1 & 0 & 1\\
      0 & 0 & 1 & -1\\
      -2 & 1 & 1 & 2\\
    \end{tabular}$\leq$ 1
\caption{Facet of 33+1, for which the quantum bound is $\sqrt{2}-1$, for a local bound of zero and a 1-bit bound of one. When intersected with the NS space, this inequality reduces to a sum of CHSH inequalities.}
\label{table7}
\end{table}

This inequality corresponds, in the NS subspace, to a sum of one CHSH inequality which uses Alice's inputs $x=0,2$ and Bob's inputs $y=0,2$ and another CHSH which uses $x=1,2$ and $y=1,2$. Therefore, the quantum bound of this inequality is $\sqrt{2}-1\approx 0.4142$, which is two times the amount of violation for CHSH. The quantum bound is obtained for the maximally entangled state $1/\sqrt{2}(|00\rangle +|11\rangle)$.

One can also have a single $I_{3322}$ contained in the facet, as the following example shows (inequality number 529)

\begin{table}[!htbp]
\begin{tabular}[c]{|c|ccc|}
\hline
-1&0&1&1\\
  &0&0&-1\\
  \hline
0&1&1&-1\\
  &-1&-1&1\\
  \hline
-1& 1&-1&1\\
  &0&0&0\\
\hline
\end{tabular}$\ \leq$ 1
$\ \xrightarrow{\mathrm{NS}} \ $
\begin{tabular}{l|c c c}
       & -1 & -1 & 0 \\
      \hline
      -1 & 0 & 1 & 1\\
      0 & 1 & 1 & -1\\
      -1 & 1 & -1 & 1\\
    \end{tabular}$\leq$ 1
\caption{Facet of 33+1, for which the quantum bound is $0.25$, for a local bound of zero and a 1-bit bound of one. When intersected with the NS space, this inequality reduces to $I_{3322}$. In fact, we see that it corresponds to $I_{3322}^{\mathrm{sym}}$ if we permute Alice's inputs $x=1$ and $x=2$. This inequality is maximally violated by the maximally entangled state and it's quantum bound is the $I_{3322}$ quantum bound.}
\end{table}

Next, we give an example of a facet which when intersected with the NS space reduces to a CHSH inequality and some other terms. Despite the extra terms, its quantum bound is the maximum violation of CHSH, attained for the maximally entangled state. This inequality is number 380 in Table X. This inequality has a similarity with the inequality of Table \ref{table6}, in the sense that both are constructed from Bell inequalities and have in addition some extra terms which do not contribute to the quantum bound. If we would remove these extra terms, the quantum and local bounds would therefore not change. Understanding how these extra terms arise could lead to a better understanding of how to construct Bell+1 inequalities from Bell inequalities.

\begin{table}[!htbp]
\begin{tabular}[c]{|c|ccc|}
\hline
0&-1&1&0\\
  &0&-1&1\\
  \hline
-2&0&2&2\\
  &1&-1&-1\\
  \hline
-1& 1&1&0\\
  &-1&0&-1\\
\hline
\end{tabular}$\ \leq$ 1
$\ \xrightarrow{\mathrm{NS}} \ $
\begin{tabular}{l|c c c}
       & 0 & -2 & -1 \\
      \hline
      0 & -1 & 1 & 0\\
      -2 & 0 & 2 & 2\\
      -1 & 1 & 1 & 0\\
    \end{tabular}$\leq$ 1
$\ = \ $
\begin{tabular}{l|c c c}
       & 0 & -1 & 0 \\
      \hline
      0 & -1 & 1 & 0\\
      0 & 0 & 0 & 0\\
      -1 & 1 & 1 & 0\\
    \end{tabular}$+$
\begin{tabular}{l|c c c}
       & 0 & -1 & -1 \\
      \hline
      0 & 0 & 0 & 0\\
      -2 & 0 & 2 & 2\\
      0 & 0 & 0 & 0\\
    \end{tabular}$\leq$ 1
\caption{Facet of 33+1, for which the quantum bound is $1/2(\sqrt{2}-1)$, for a local bound of zero and a 1-bit bound of one. When intersected with the NS space, this inequality reduces to a CHSH inequality for two of each party's inputs and some other terms. This inequality is maximally violated by the maximally entangled state and it's quantum bound is the CHSH quantum bound.}
\label{table7}
\end{table}

Most inequalities of 33+1 have a quantum bound which is different than the bound of CHSH, $I_{3322}$ or twice their amount. Most inequalities have quantum bounds that do not relate easily to Bell inequalities for binary outcomes, up to three settings. As a final example, we show such an 33+1 facet, and how its NS intersection is constructed from CHSH and $I_{3322}$ even if the quantum bound does not relate directly to the maximal violations of the Bell inequalities. One such inequality is inequality number 196 in Table X,

\begin{table}[!htbp]
\begin{tabular}[c]{|c|ccc|}
\hline
-1&0&1&1\\
  &0&0&0\\
  \hline
-2&2&-1&2\\
  &-1&0&-1\\
  \hline
0&2&1&-2\\
  &-1&-1&1\\
\hline
\end{tabular}$\ \leq$ 1
$\ \xrightarrow{\mathrm{NS}} \ $
\begin{tabular}{l|c c c}
       & -2 & -1 & 0 \\
      \hline
      -1 & 0 & 1 & 1\\
      -2 & 2 & -1 & 2\\
      0 & 2 & 1 & -2\\
    \end{tabular}$\leq$ 1
$\ = \ $
\begin{tabular}{l|c c c}
       & -1 & -1 & 0 \\
      \hline
      -1 & 0 & 1 & 1\\
      -1 & 1 & -1 & 1\\
      0 & 1 & 1 & -1\\
    \end{tabular}$+$
\begin{tabular}{l|c c c}
       & -1 & 0 & 0 \\
      \hline
      0 & 0 & 0 & 0\\
      -1 & 1 & 0 & 1\\
      0 & 1 & 0 & -1\\
    \end{tabular}$\leq$ 1
\caption{Facet of 33+1, for which the quantum bound is $0.4158$, for a local bound of zero and a 1-bit bound of one. When intersected with the NS space, this inequality reduces to a sum of a CHSH inequality for two of each party's inputs and an $I_{3322}$. This inequality is maximally violated by the non-maximally entangled state.}
\label{table8}
\end{table}

As shown in Table \ref{table8}, facet number 196 corresponds to a sum of $I_{3322}^{\mathrm{sym}}$ and CHSH using inputs $x=1,2$ of Alice and $y=0,2$ of Bob. maximum violation of 0.4158 is given by a partially entangled state
\begin{equation}
|\psi\rangle=0.7379|00\rangle+0.6749|11\rangle
\end{equation} 
and the resistance to noise for this inequality is $\lambda =0.7830$, larger than the resistance to noise of CHSH ($\lambda_{\mathrm{CHSH}}=0.7071$) but lower than the resistance to noise of $I_{3322}$ ($\lambda_{I_{3322}}=0.8$).

\clearpage

\includepdf[pages={{},1,{},2,{},3,{},4,{},5,{},6,{},7,{},8,{},9,{},10,{},11,{},12,{},13,{},14},scale=1]{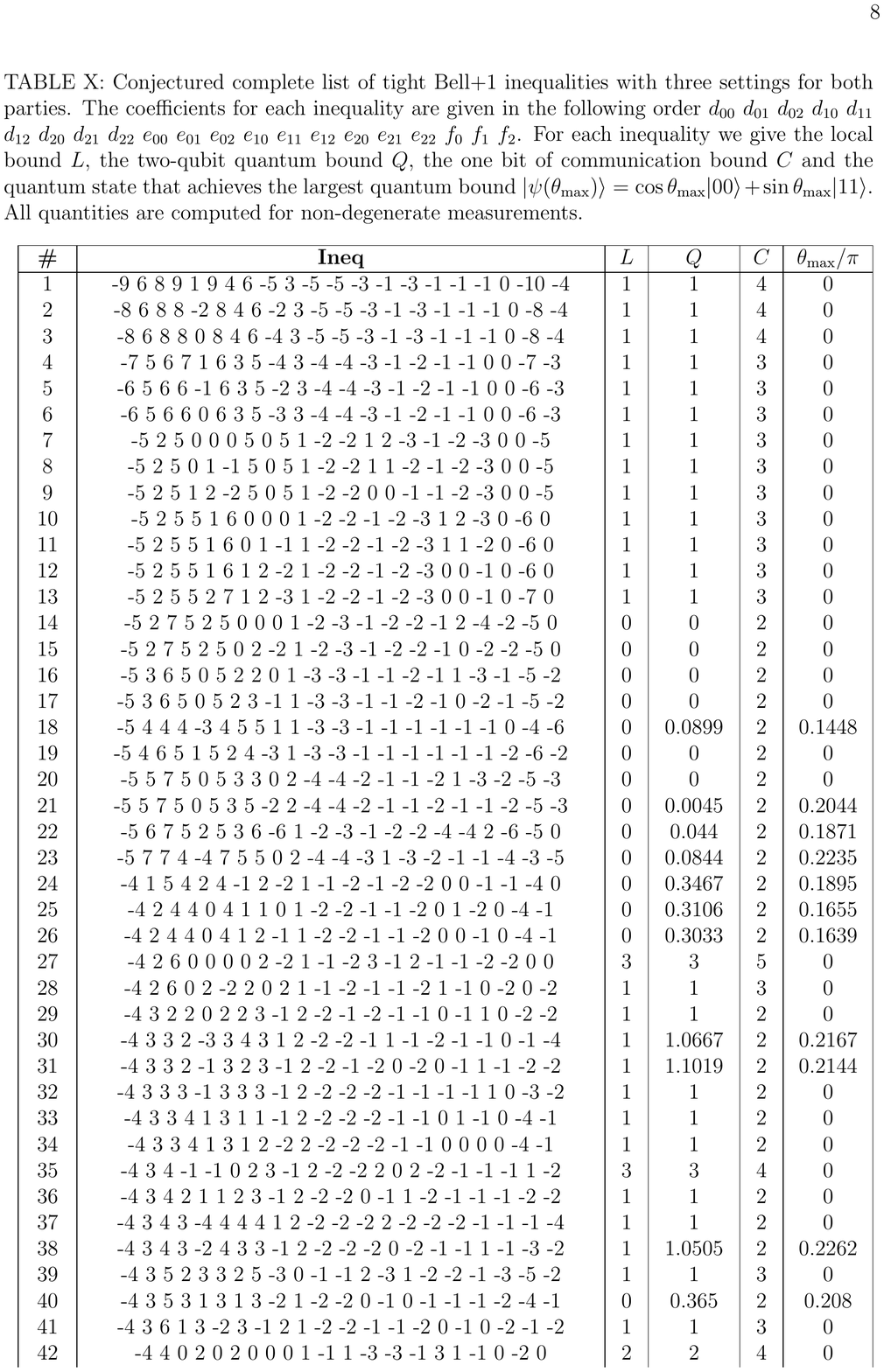}

\end{document}